\documentclass[a4paper, 12pt]{iopart}
\usepackage{graphicx}
\usepackage{amssymb}
\usepackage{hyperref}
\usepackage{doi}
\usepackage{braket}
\usepackage{bm}
\newcommand{\bea}{\begin{eqnarray}}
\newcommand{\eea}{\end{eqnarray}}
\newcommand{\beq}{\begin{equation}}
\newcommand{\eeq}{\end{equation}}

\newcommand{\up}{\uparrow}
\newcommand{\down}{\downarrow}

\begin{document}
\title{Time-dependent spectral properties of photoexcited one-dimensional ionic Hubbard model: an exact diagonalization study}

\author{Junichi Okamoto}
\address{Physikalisches Institut, Albert-Ludwigs-Universit\"at Freiburg, Hermann-Herder-Stra\ss e 3, 79104 Freiburg, Germany}

\ead{junichi.okamoto@physik.uni-freiburg.de}

\date{\today}

\begin{abstract}
Motivated by the recent progress in time-resolved nonequilibrium spectroscopy in condensed matter, we study an optically excited one-dimensional ionic Hubbard model by exact diagonalization. The model is relevant to organic crystals, transition metal oxides, or ultracold atoms in optical lattices. We implement numerical pump-probe measurements to calculate time-dependent conductivity and single-particle spectral functions. In general, short optical excitation induces a metallic behavior imprinted as a Drude peak in conductivity or an in-gap density of states. In a Mott insulator, we find that the induced Drude peak oscillates at the pump frequency and its second harmonic. The former comes from the oscillation of currents, and the latter from the interference of single- and three-photon excited states. In a band insulator, the Drude peak oscillates only at the pump frequency, and quantities such as the double occupancy do not oscillate. The absence of the second harmonic oscillation is due to the degeneracy of multi-photon excited states. The in-gap density of states in both insulators correlates with the Drude weight and the energy absorption for weak pumping. Strong pumping leads to saturation of the in-gap density of states and to suppression of the Drude weight in the Mott regime. We have also checked that the above features are robust for insulators in the intermediate parameter range. Our study demonstrates the distinct natures of the multi-photon excited states in two different insulators.
  
\end{abstract}

\section{Introduction}
Nonequilibrium properties of strongly correlated electron systems have been an intense research topic for many decades \cite{pitaevskii1981, carmelo2007, haug2008, yamada2010, avella2011, avella2013, avella2016}. Recently systems under strong optical excitation attract considerable interest due to the development of intense laser fields \cite{averitt2002, yonemitsu2008, giannetti2016, nicoletti2016, citro2018, ishihara2019}. Resultant states after optical excitation are far away from equilibrium, and hence various time-dependent spectroscopic measurements are employed to analyze the transient states, e.g., time-domain THz spectroscopy \cite{schmuttenmaer2004}, time-resolved photoemission spectroscopy \cite{zhou2018, lv2019}, or time-dependent X-ray diffraction \cite{wark1996, buzzimichele2019}. They have been applied to systems such as high-$T_c$ superconductors \cite{giannetti2016}, manganites \cite{averitt2002}, or organic compounds \cite{yonemitsu2008}, and have discovered intriguing phenomena such as light-induced superconductivity \cite{fausti2011,hu2014,mitrano2016}, metastable charge-order \cite{stojchevska2014}, or ultrafast structural transitions in manganites \cite{rini2007, tobey2008}.

Theoretically it is challenging to understand these experimental results due to the strong correlation of the systems and nonequilibrium nature of the problem. Various static methods have been generalized to time-domain to describe the highly nontrivial dynamics, e.g., dynamical mean-field theory (DMFT) \cite{Aoki2014}, density-matrix renormalization group (DMRG) \cite{cazalilla2002, vidal2004, daley2004, white2004, garcia-ripoll2006, paeckel2019}, Gutzwiller approximations \cite{schiro2010, fabrizio2013}, Monte Carlo methods \cite{goth2012, ido2015}, and exact diagonalization \cite{park1986, manmana2005, balzer2012}. In particular, DMFT has been widely used for Hubbard-like models, and been very successful to account for recent results of pump-probe spectroscopy \cite{eckstein2009, giannetti2016, citro2018}. Also, DMRG gives very accurate results for one-dimensional systems, and extensions to higher-dimensions are possible \cite{Orus2014}.

In this work, we study an ionic Hubbard model, which can be considered as an effective model of a Holstein-Hubbard model where frozen phonons are dimerized to give static staggered potential modulation. Alternatively, we can regard it as a variant of Falicov-Kimball model \cite{falicov1969} where $f$-electrons are charge-ordered and frozen. Here we ignore the dynamics of the underlying objects that give the staggered modulation. This means that our results are relevant to experimental data for a short period of time before the ionic motion sets in; the model serves as a starting point for further investigation with dynamical phonons. There is also an experimental realization of the model in ultracold atoms \cite{atala2013}.

We employ an exact diagonalization method to solve the time-dependent Schr\"odinger equation, and implement numerical pump-probe measurements to obtain time-dependent conductivity \cite{lenarcic2014, lu2015, shao2016} and single-particle spectral functions \cite{wang2017, wang2018}. Short optical pulses applied to a Mott insulator or a band insulator make the system metallic, which is indicated by a nonzero Drude peak in conductivity. We find that the Drude peak oscillates at the pump frequency in both cases, while also at the second harmonic in the Mott insulator \footnote{We note that this is nothing to do with the second harmonic generation.}. The second harmonic oscillation in the Mott regime is explained by a constructive interference of single- and three-photon excited states. We have confirmed that the distinct dynamic behaviors still appear for intermediate cases unless the energy gap becomes too small. The relation between the in-gap density of states and the Drude weight is also discussed.

The paper is organized as follows. In \sref{sec:model}, an ionic Hubbard model is introduced. \Sref{sec:method} is devoted to the details of the exact diagonalization method and pump-probe protocols. In \sref{sec:results}, we show the results for Mott and band insulators.  The intermediate cases are also discussed. In \sref{sec:diss}, we reveal the connection between the distinct energy-level structures in the two insulators and the oscillations in observables. In \sref{sec:conc}, we conclude. The details of calculations and additional results not covered in the main texts are summarized in appendices.

\section{Model}
\label{sec:model}
We consider a one-dimensional ionic Hubbard model
\begin{equation}
H = \sum_{l} - J_0\left(\sum_{\sigma}c^{\dagger}_{l \sigma} c_{l+1, \sigma} + \rm{h.c.} \right) + \Delta (-1)^l n_l  + U n_{l \up} n_{l \down}, 
\label{eq:Hamiltonian}
\end{equation}
where $J_0$ is the hopping amplitude, $c_{l \sigma}$ and $c^{\dagger}_{l \sigma}$ annihilation and creation operators at site $l$ of spin $\sigma$, $\Delta$ the staggered potential, and $U$ the on-site Hubbard interaction. We set $J_0 = 1$ in the remainder of the paper. The density operators are given by $n_{l \sigma} = c^{\dagger}_{l \sigma} c_{l \sigma}$ and $n_{l} = n_{l \up} + n_{l \down}$. We focus on half-filling, where the number of particles $N = \sum_l n_l$ is equal to the number of sites $L$. This model has been investigated to understand organic crystals or transition metal oxides \cite{nagaosa1986, egami1993, ishihara1994, maeshima2005}. It has also an intriguing topological property related to the Zak phase \cite{resta1995, valiente2010}, which can be measured experimentally using ultracold atoms in an optical lattice \cite{atala2013}.

When the Hubbard interaction $U$ is dominant ($U\gg J_0,\Delta$), the system is a Mott insulator (MI) with a spin-density wave or antiferromagnetic ordering. When the Hubbard interaction $U$ is negligible, the staggered potential $\Delta$ leads to a band dispersion
\begin{equation}
\epsilon^{\pm}(k) = \pm \sqrt{4 J_0^2 \cos^2 (k) + \Delta^2},
\label{eq:BI_band}
\end{equation}
where a gap of $2\Delta$ opens at $k=\pi/2$. For a half-filled system, the lower band is completely occupied, and the upper band is empty in the ground state. Therefore, it is a band insulator (BI).

In this work, we describe optical fields by a spatially uniform time-dependent vector potential $A(t)$, and include it by Peierls' substitution, $J_0 \rightarrow J_0 \exp [ i A(t) ]$ (we use $e = \hbar = c = 1$); the hopping amplitude becomes complex. To investigate nonequilibrium dynamics under optical fields, we employ pump and probe pulses. For pump pulses, we consider the following Gaussian pulse
\begin{equation}
A_{\rm pump} (t) = A_0 \exp \left[ - 4 \log(2)(t-t_0)^2/\tau^2 \right] \sin(\omega_p (t-t_0)) ,
\label{eq:A_pump}
\end{equation}
where $A_0$ is the pump strength, $t_0$ the central pump time, and $\tau$ the full width at half maximum, and $\omega_p$ the pump frequency. As a probe pulse (if included in the simulations), we use a step function at time $t_p$
\begin{equation}
A_{\rm probe} (t, t_p) = -E_0 \Theta(t - t_p),
\label{eq:probe}
\end{equation}
resulting in an electric field of a delta function $E_{\rm probe} = E_0 \delta(t - t_p) $. This instantaneous form is taken for simplicity instead of narrow Gaussian forms, which are more realistic for experiments. The current density is given by
\begin{equation}
j = - \sum_{l, \sigma} \left[ i J_0 e^{i A(t)} c^{\dagger}_{l \sigma} c_{l+1, \sigma} + {\rm h.c.} \right]/L.
\label{current_op}
\end{equation}

\section{Method}
\label{sec:method}
We use an exact diagonalization method to study the ionic Hubbard model in \eref{eq:Hamiltonian}. A ground state is found by the conventional Lanczos method \cite{lanczos1950}, and then used as an initial state for the time-dependent Schr\"odinger equation. Time evolution is implemented by the Krylov subspace method based on the Lanczos method \cite{park1986, manmana2005, balzer2012}. We use $\tau = 5/J_0$ and match the pump frequency $\omega_p$ to the first peak of the linear absorption spectrum unless otherwise explicitly stated. The probe pulse strength $E_0$ is set to $0.005$; we have checked that the results are independent of the probe strength for $E_0 \leq 0.01$. $L=N=10$ with the periodic boundary condition is used. 

In the following sections, we investigate time-resolved transient spectral properties of photoexcited states in the ionic Hubbard model by two protocols. One is time-resolved transient conductivity given by two-dimensional pump-probe analysis. The other is a transient single-particle spectral function, which can be measured by time-resolved photoemission experiments. In the following, we briefly explain the simulation setups for the two protocols.  
 
\subsection{Time-resolved transient conductivity $\sigma(t, \omega)$}
In order to obtain time-resolved transient conductivity, we follow the formulation of Ref. \cite{kindt1999} (see also Refs. \cite{nemec2002, lenarcic2014, lu2015, orenstein2015, shao2016, kennes2017, okamoto2017}.) A general conductivity response function $\Sigma (t, t')$ under a weak probe field $E_{\rm probe}$ is defined by \cite{vengurlekar1988}
\begin{equation}
\delta j(t) = \int_{-\infty}^{t} \Sigma(t, t') E_{\rm probe} (t') dt'
\end{equation}
Here $\delta j (t)$ is a current solely induced by the probe field. When currents are nonzero even without a probe field (e.g. due to a pump field), such contributions are subtracted to obtain $\delta j(t)$. Due to causality, $\Sigma(t, t') =0$ for $t < t'$. In equilibrium $\Sigma(t,t')$ only depends on the time difference $t-t'$ because of the temporal translation symmetry.

We consider a weak impulsive probe to the system at $t_p$ as \eref{eq:probe}. In real pump-probe experiments, a probe field has a finite duration. If its duration is short enough, the results are essentially the same as the one with the impulsive probe \cite{shao2016}. Repeating the pump-probe simulations with different time delay $t_p$, we obtain the two-dimensional current responses $\delta j(t, t_p)$ in time domain
\begin{equation}
\delta j(t, t_p) = \int_{-\infty}^{t} \Sigma(t, t') E_{\rm probe} (t', t_p) dt' = \Sigma(t, t_p) E_0.
\end{equation}
Changing one of the time variables to $s \equiv t - t_p$, we introduce  
\begin{equation}
\tilde{\delta j} (t, s = t- t_p) \equiv \delta j (t, t_p), \ \ \tilde{\Sigma} (t, s= t- t_p) \equiv \Sigma (t, t_p).
\end{equation}
This leads to 
\begin{equation}
\tilde{\delta j} (t, s) = \tilde{\Sigma} (t, s) E_0.
\end{equation}
Finally Fourier transforming over $s$, we find
\begin{equation}
\sigma(t, \omega) \equiv \int ds \tilde{\Sigma}(t, s) e^{-i \omega s} = \tilde{\delta j} (t, \omega)/E_{0} .
\label{eq:sigma}
\end{equation}
In practice, we use a small damping factor $\eta = 1/L$ in the Fourier transformation to remove artifacts of the finite window.

\subsection{Time-dependent transient single-particle spectral function $A(t,\omega)$}
In photoemission experiments, the central quantity of interest is the probability to detect an electron kicked out by photons. Assuming a uniform strength of probing laser pulses for duration of $2T$, the probability of detecting an ejected electron after the probe pulse is expressed by lesser and greater Green's functions as \cite{freericks2009, braun2015, sentef2017, wang2017, wang2018, acciai2019}
\begin{eqnarray}
A^{\alpha}(t, \omega) = A^{\alpha }_{\rm PES} (t, \omega) + A^{\alpha }_{\rm IPES} (t, \omega), \label{eq:A_formula} \\
A^{\alpha}_{\rm PES} (t, \omega) = {\rm Re} \int_{-T}^{T} d \tau_{1} \int_{-T}^{T} d \tau_{2}e^{-i \omega\left(\tau_{1}-\tau_{2}\right)} \langle c_{\alpha}^{\dagger}\left(t+\tau_{1}\right)  c_{\alpha}\left(t+\tau_{2}\right)\rangle, \label{eq:A_PES} \\
A^{\alpha}_{\rm IPES} (t, \omega) = {\rm Re} \int_{-T}^{T} d \tau_{1} \int_{-T}^{T} d \tau_{2}e^{-i \omega\left(\tau_{1}-\tau_{2}\right)} \langle c_{\alpha}\left(t+\tau_{2}\right) c_{\alpha}^{\dagger}\left(t+\tau_{1}\right)\rangle, \label{eq:A_IPES}
\end{eqnarray}
where we have assumed that the coupling between incoming photons and electrons does not depend on their states. The first term corresponds to the photoemission spectroscopy (PES), and the second one to the inverse photoemission spectroscopy (IPES). When the Hamiltonian is time-independent, the expression is reduced to the conventional formula in the limit $T \rightarrow \infty$ \cite{freericks2009, braun2015}. Here, instead, the time integral is limited by width $2T$, which naturally gives a broadening of a spectral peak $\sim \pi/T$. We use $T=10$ in the following calculations.
\begin{figure*}[t!]\center
\includegraphics[scale=0.8]{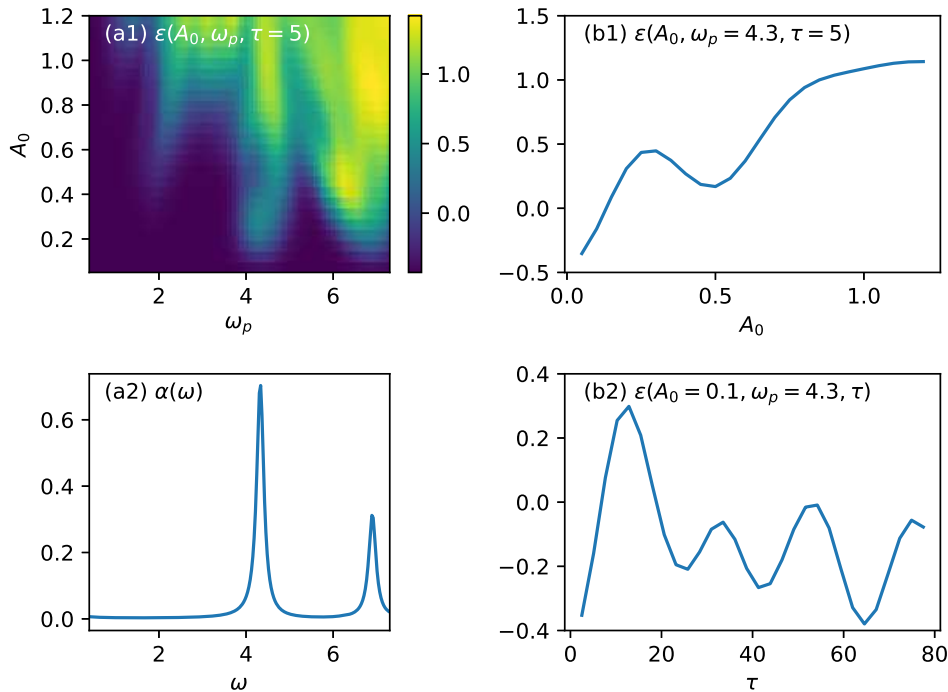}
\caption{(a) The energy of the MI after photoexcitation of amplitude $A_0$ and frequency $\omega_p$. The optical excitation is allowed at $\omega_p \approx 4.3$ and $6.9$. (b) The absorption spectrum of the MI. (c) The line cut of plot (a) at $\omega_p = 4.3$. (d) $\tau$ dependence of the energy.}
\label{fig:MI_scan}
\end{figure*}

\section{Results}
\label{sec:results}
\subsection{Mott insulators}
\label{subsec:MI}
We first look at a Mott insulator, $U=6$ and $\Delta = 0$. In \fref{fig:MI_scan}(a), we plot the energy expectation value $\epsilon = \braket{H}$ after photoexcitation with pump amplitude $A_0$ and pump frequency $\omega_p$. The energy is measured with the Hamiltonian without any vector potential. We see that around $\omega_p \approx 4.3$ and $6.9$, the system absorbs energies. This is consistent with the linear absorption spectrum $\alpha(\omega)$ in \fref{fig:MI_scan}(b) calculated by \cite{dagotto1994, matsueda2007}
\begin{equation}
\alpha(\omega) = - \Im \frac{1}{\pi} \bra{\psi_0} j \frac{1}{\omega - H + \epsilon_0} j \ket{\psi_0},
\label{absp}
\end{equation}
where $\ket{\psi_0}$ is the ground state, $\epsilon_0$ the ground state energy, and $j$ the current operator in \eref{current_op}. For a fixed pump frequency $\omega_p = 4.3$ [\fref{fig:MI_scan}(c)], the energy absorption becomes nonlinear for $A_0 \gtrsim 0.3$, even with a local minimum around $A_0 \approx 0.5$. Such nonlinear dependence on $A_0$ is also observed in a charge-ordered state in an extended Hubbard model \cite{lu2012}. In \fref{fig:MI_scan}(d), we plot the energy as a function of pulse duration $\tau$. We see a large oscillation, even almost returning to the ground state around $\tau \approx 62$; this is related to the coherent nature of excited states as discussed in \Sref{sec:diss}. In the next subsection, we see that such an oscillation is more apparent in a band insulator. The double occupancy, $n_{\rm do} \equiv \sum_l \braket{\delta_{n_l, 2}}/L$ also follows the same trend (not shown) as the energy absorption, which indicates that the energy absorption occurs via the doublon-holon excitation.

\begin{figure*}[t!]\center
\includegraphics[scale=0.8]{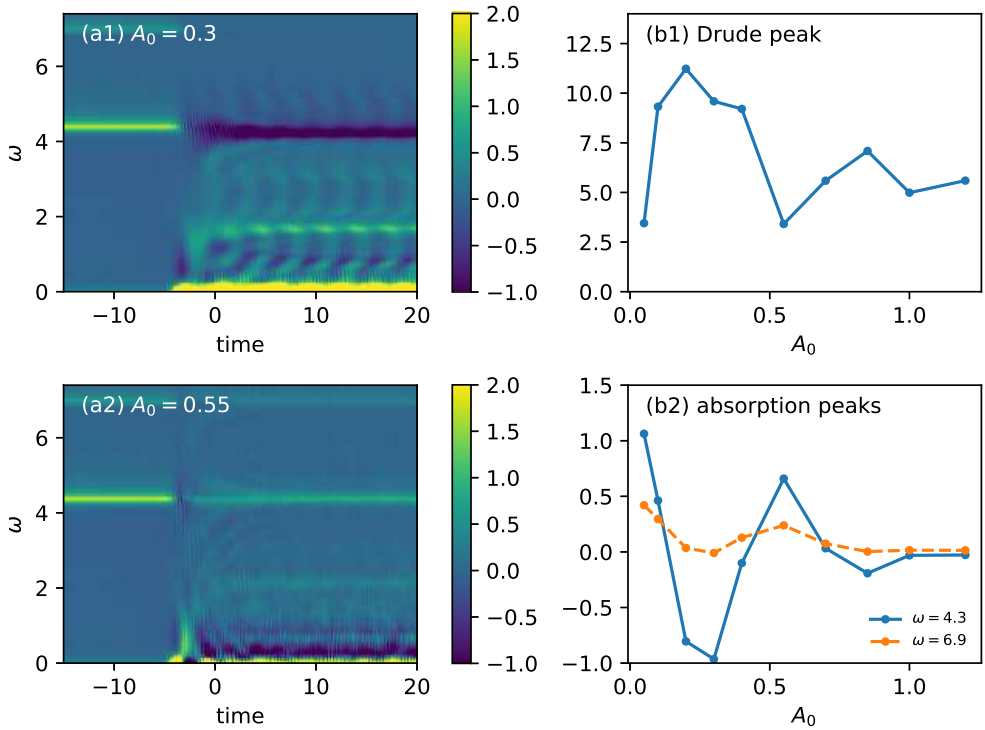}
\caption{(a) Time-dependent conductivity $\sigma(t, \omega)$ of the MI for pump strength $A_0 = 0.3$ and $0.55$. The pump pulse takes a Gaussian form \eref{eq:A_pump} with $t_0=0$ and $\tau = 5$. (b) Amplitude dependence of the Drude peak $\sigma(t, 0)$ and the absorption peaks, $\sigma(t, 4.3)$ and $\sigma(t, 6.9)$ in the MI.}
\label{fig:MI_sigma}
\end{figure*}

The real part of the time-dependent transient conductivity $\sigma(t, \omega)$ for pump strength $A_0=0.3$ and $A_0 =0.55$ is plotted in \fref{fig:MI_sigma}(a). We see that the system becomes metallic due to the optical excitation, which is clearly seen as a Drude peak at low frequencies. Furthermore, a sub gap peak around $\omega \approx 1.9$ appears, and the conductivity becomes negative slightly above the Drude peak. The sub-gap feature resembles the case of an extended Hubbard model \cite{lu2015}, which is attributed to a transition from an odd-parity state to an even-parity state. However, various other features in $\sigma(\omega)$ depend non-monotonically on the pump strength $A_0$ as the energy absorption does. In particular, visible changes occur for the three peaks at $\omega \approx 0$ (the Drude peak), $\omega \approx 4.3$ (the first absorption peak), and $\omega \approx 6.9$ (the second absorption peak), which are plotted in \fref{fig:MI_sigma}(b1) and (b2) as functions of $A_0$. For small pump amplitude $A_0 \lesssim 0.6$, the Drude peak shows a similar structure as the energy absorption, e.g., the local minimum around $A_0 \approx 0.5$. However, for stronger pump pulses $A_0 \gtrsim 0.6$, too much energy is given to the system, and the Drude peak diminishes. The second peak at $\omega \approx 6.9$ follows a similar trend. On the other hand, due to the resonant condition, the first excited peak at $\omega \approx 4.3$ is depleted even to negative values when the energy absorption is small \cite{maeshima2005a}, and for larger $A_0\gtrsim 0.7$, the peak disappears.

\begin{figure*}[t!]\center
\includegraphics[scale=0.8]{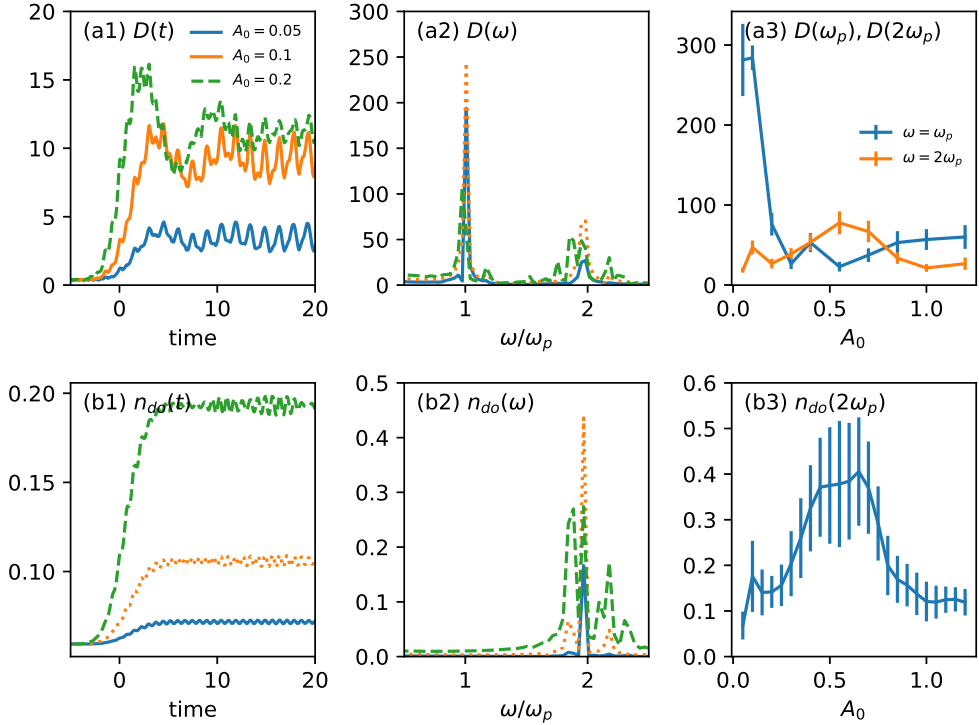}
\caption{(a) Time evolution, the Fourier spectrum, and the amplitude dependence of the Fourier components of the Drude peak, $D$, in the MI. (b) The same quantities for the double occupancy $n_{\rm do}$.}
\label{fig:MI_osci}
\end{figure*}

We also observe that the conductivity oscillates in time, which is most evident for the Drude peak. In \fref{fig:MI_osci}(a1) and (a2), the time evolution of the Drude peak $D(t) = \sigma(t,0)$ and its Fourier spectrum, $D(\omega)$, are plotted. This shows that the oscillation has exactly the pump frequency $\omega_p$ and its second harmonic $2\omega_p$. The fundamental mode comes from the oscillation of the current at $\omega_p$. On the other hand, the double occupancy $n_{\rm do}$ oscillates only at $2\omega_p$ [\fref{fig:MI_osci}(b1) and (b2)]. Similarly, other quantities such as the kinetic energy and antiferromagnetic order oscillate at the second harmonic only. As we show in the next section using full diagonalization of a smaller system, such second harmonic oscillations come from the interference of the multi-photon excited states. The peaks in the Fourier spectra of the Drude weight and of the double occupancy are fitted by a simple Gaussian, and their $A_0$ dependence is plotted in \fref{fig:MI_osci}(a3) and (b3). The second harmonic components, $D(2\omega_p)$ and $n_{\rm do} (2 \omega_p)$, both have maximum around $A_0 \approx 0.6$, and follow the same trend; a part of the Drude peak is governed by the doublon-holon excitation. On the other hand, the fundamental component $D(\omega_p)$ decreases significantly once the pump strength $A_0$ gets strong. 
  
\begin{figure*}[t!]\center
\includegraphics[scale=0.8]{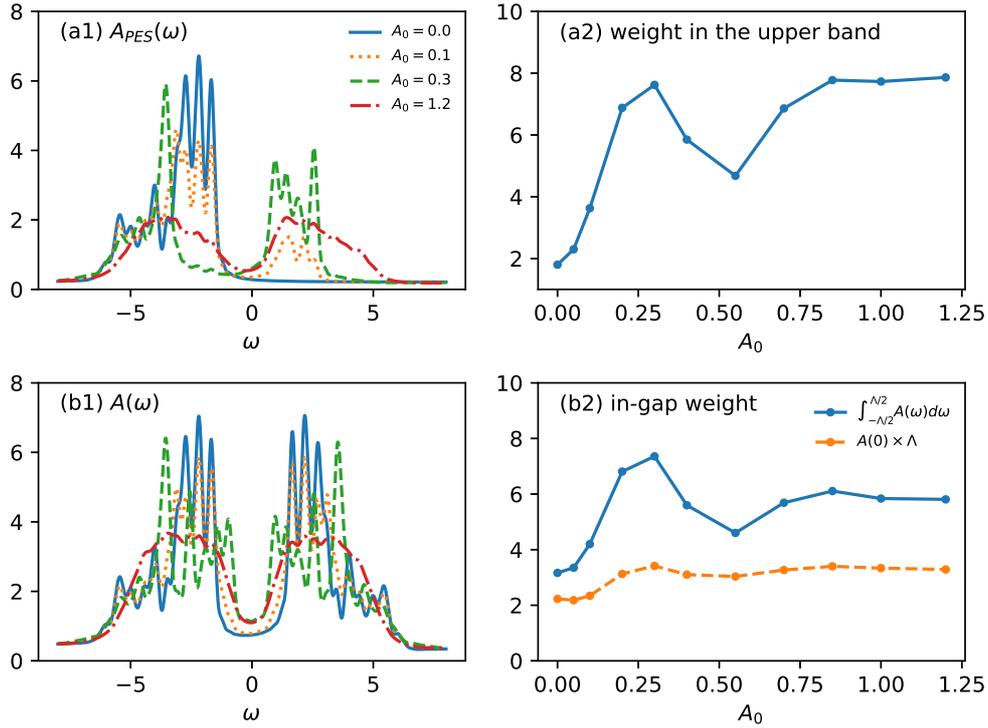}
\caption{(a1) PES signals $A_{\rm PES}(t_f, \omega)$ and (b1) single-particle spectral functions $A(t_f, \omega)$ of the MI for various pump strength $A_0 = 0, 0.1, 0.3$ and $1.2$. (a2) The integrated density of states for the upper band. (b2) The integrated in-gap density of states (solid line) and $A(t_f, 0)$. We use $\Lambda = 3.0$ as the size of the gap.}
\label{fig:MI_A}
\end{figure*}

Now we turn to the single-particle spectral functions $A (t, \omega)$ for a single-particle state $\alpha = (l=1,\uparrow)$. We focus on a final state at $t_f = t_0 + 50$, which is long after the pump pulse at $t_0$. In \ref{sec:AppA}, we show some other transient cases $|t - t_0| < 50$. First, \fref{fig:MI_A}(a1) shows the PES signal $A_{\rm PES}(\omega)$ for various $A_0$. The spectral weight is transferred from the lower Hubbard band to the upper Hubbard band, and the transferred weight as a function of $A_0$ is plotted in \fref{fig:MI_A}(a2). It shows a similar $A_0$ dependence as the energy absorption in \fref{fig:MI_scan}. Interestingly, around $A_0 \approx 0.3$, the weight is widely depleted in $-3.5 \lesssim \omega \lesssim 1.5$ including the original Hubbard gap, in contrast to the case of $A_0 = 1.2$ where only the slightly filled Hubbard gap remains. In \fref{fig:MI_A}(b1), we plot $A(t_f, \omega)$ for various $A_0$. We see that the gap is gradually filled with increasing $A_0$, while the upper and lower Hubbard bands do not completely merge. In \fref{fig:MI_A}(b2), we show the integrated in-gap density of states $\int_{-\Lambda/2}^{\Lambda/2} A(t_f, \omega) d \omega$ with $\Lambda = 3.0$ and $A(t_f, 0)\Lambda$ as functions of $A_0$. This again shows a similar behavior as the energy absorption. Therefore, we conclude that the in-gap density of states represents photoexcited carriers, which do not necessarily contribute to the coherent Drude peak, especially at a large pump strength $A_0$.

\subsection{Band insulators}
\label{subsec:BI}

\begin{figure*}[t!]\center
\includegraphics[scale=0.8]{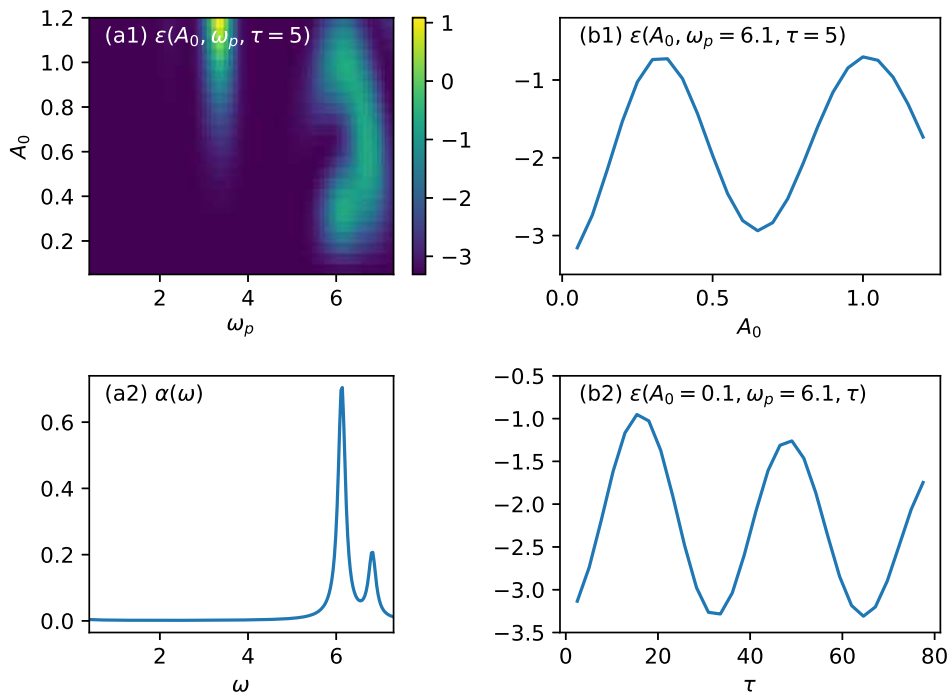}
\caption{(a) The energy of the BI after photoexcitation of amplitude $A_0$ and frequency $\omega_p$. The optical excitation is allowed at $\omega_p \approx 6.1$ and $6.9$. (b) The absorption spectrum of the BI. (c) The line cut of plot (a) at $\omega_p = 6.1$. (d) $\tau$ dependence of the energy.}
\label{fig:BI_scan}
\end{figure*}

As a band insulator, we take $\Delta = 3$, and $U=0$. \Fref{fig:BI_scan}(a) shows the energy as a function of pump strength $A_0$ and frequency $\omega_p$. The optical excitation occurs at $\omega \approx 6.1$ and $6.9$ as shown in \fref{fig:BI_scan}(b). As in the MI, the energy absorption depends non-monotonically on $A_0$ at a fixed $\omega_{p}$. We also find that the antiferromagnetic order is enhanced near the first excitation frequency, which is shown in \ref{sec:AppB}. We expect that the spin part does not affect the charge dynamics that we are interested in due to charge-spin separation in one dimension, and thus we do not elaborate on it here. \Fref{fig:BI_scan}(c) and (d) shows the $A_0$ and $\tau$ dependence of the energy. We see nearly perfect oscillations both in terms of $A_0$ and $\tau$. Since the system has two bands separated by $2\Delta$, which resembles a two-level system, we consider these oscillations as Rabi-like oscillations. In the case of the MI, correlation effects hinder the simple two-level analogy (see \fref{fig:full_eig}), and deviations from a perfect oscillation occur especially at strong pump strength.

\begin{figure*}[t!]\center
\includegraphics[scale=0.8]{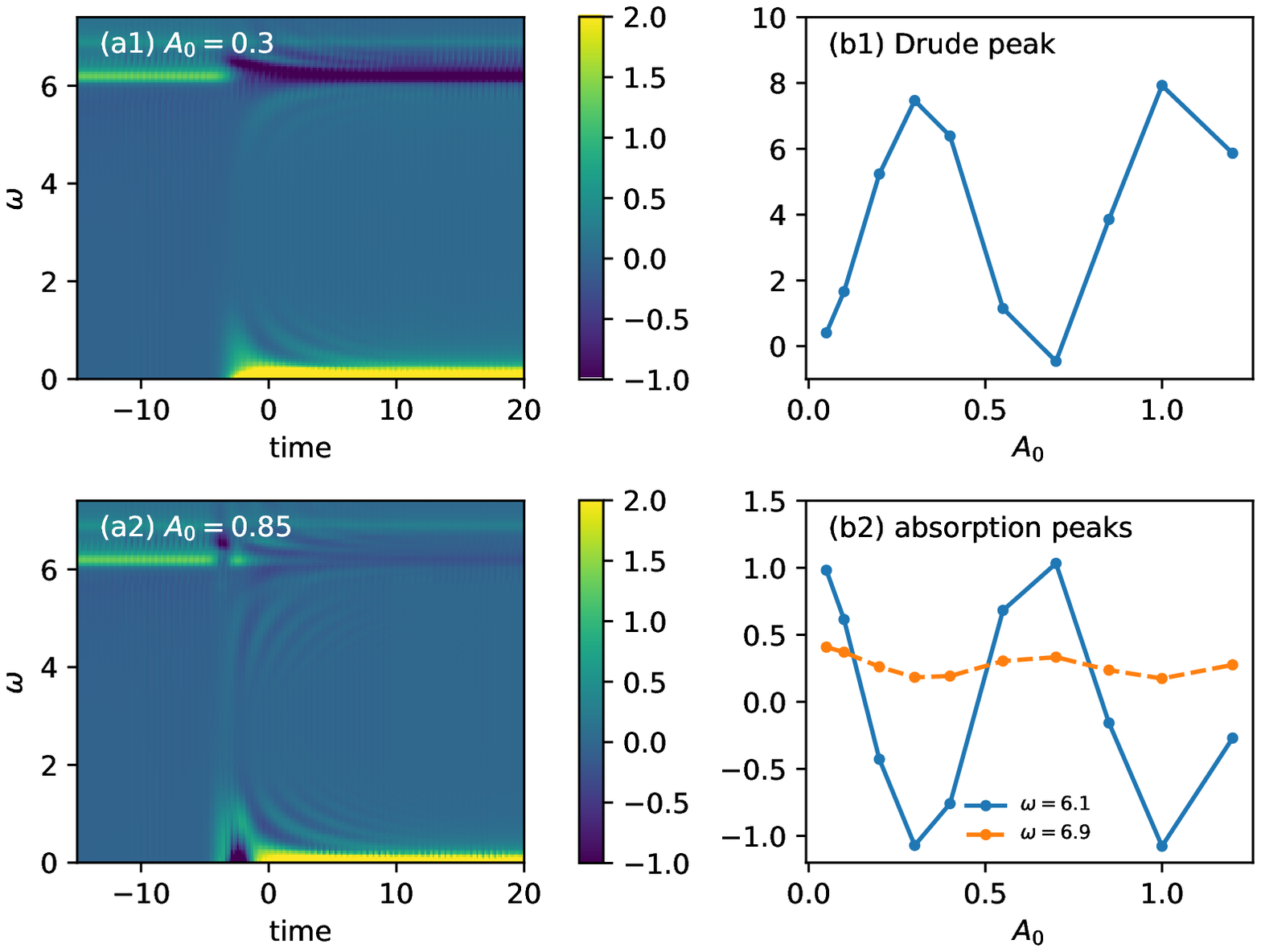}
\caption{(a) Time-dependent conductivity $\sigma(t, \omega)$ of the BI for pump strength $A_0 = 0.3$ and $0.85$. The pump pulse takes a Gaussian form \eref{eq:A_pump} with $t_0=0$ and $\tau = 5$. (b) Amplitude dependence of the Drude peak $\sigma(t, 0)$ and the absorption peaks, $\sigma(t, 6.1)$ and $\sigma(t, 6.9)$ in the BI.}
\label{fig:BI_sigma}
\end{figure*}

The time-dependent transient conductivity $\sigma(t, \omega)$ for various pump strength $A_0$ is plotted in \fref{fig:BI_sigma}(a). Contrary to the MI, there is no sub gap peak below the first excited peak $\omega \approx 6.1$, nor the negative conductivity slightly above the Drude peak. Therefore, these features appearing in the MI are supposed to originate from the correlation effects \cite{lu2015}. The non-monotonic dependence of the conductivity peaks on $A_0$ is shown in \fref{fig:BI_sigma}(b). Up to the pump strength we studied $A_0 \leq 1.2$, we see that the Drude peak correlates with the energy absorption in \fref{fig:BI_scan}. This is different from the MI, where the Drude peak does not follow the energy absorption for large $A_0$. We expect that in the case of the BI the photodoped carriers independently contribute to the Drude peak, since there is no interaction. However, in the case of the MI, the photodoped carriers need to form coherent quasi particles to contribute to the Drude peak, which becomes difficult when the system absorbs too much energy. The peak at the pump frequency $\omega_p$ changes signs due to the resonant condition, while the peak amplitude does not decay as increasing $A_0$.

\begin{figure*}[t!]\center
\includegraphics[scale=0.8]{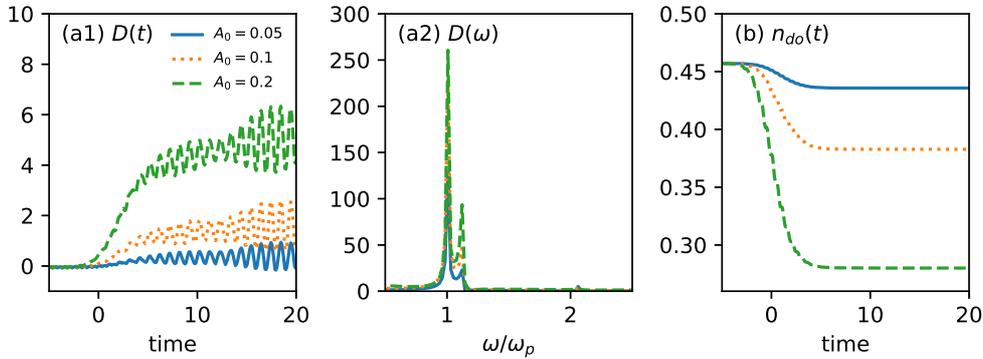}
\caption{(a) Time evolution and the Fourier spectrum of the Drude peak, $D$, in the BI. (b) Time evolution of the double occupancy $n_{\rm do}$.}
\label{fig:BI_osci}
\end{figure*}

The oscillating features in $\sigma(t, \omega)$ in the BI are also different from the ones of the MI. In \fref{fig:BI_osci}(a), we plot the time evolution and the Fourier spectrum of the Drude peak, $D$. We find that the Drude peak oscillates at the pump frequency $\omega_p = 6.1$, which comes from the current oscillation. However, there is no oscillation at the second harmonic as in the MI. Similarly, the double occupancy plotted in \fref{fig:BI_osci}(b) shows no oscillation. The lack of the second harmonic oscillations in the BI is an important difference compared to the MI. As we discuss in more detail in \sref{sec:diss}, the absence is due to the structure of the excitation spectrum of the BI. In the BI, many multi-photon excited states exist near multiples of the energy gap $2\Delta$. Such degeneracy and the incoherent phases destroy the constructive interference. 

\begin{figure*}[t!]\center
\includegraphics[scale=0.8]{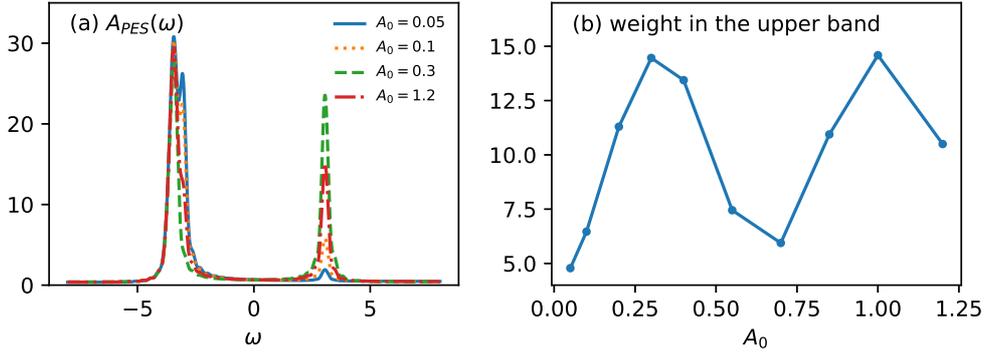}
\caption{(a) PES signals $A_{\rm PES}(t_f, \omega)$ of the BI for various pump strength $A_0 = 0.05, 0.1, 0.3$ and $1.2$. (b) The integrated density of states for the upper band.}
\label{fig:BI_A}
\end{figure*}

Before we continue for the full-diagonalization results, we briefly comment on the single-particle spectral function $A(t_f, \omega)$ for the BI. Here due to the doubling of the unit cell, we take the average of $A^{\alpha}$ with $\alpha = (l=1, \uparrow)$ and $(l=2, \uparrow)$. Furthermore, in the non-interacting limit, the sum of the photoemission and inverse photoemission in \eref{eq:A_formula} is independent of the initial states (see \ref{sec:AppC}). Therefore, we consider only the contribution from the photoemission part $A^{\alpha}_{\rm PES}$. \Fref{fig:BI_A}(a) shows the PES signals for different pump strength $A_0$. We see that the upper band is more populated as the pump strength becomes larger. The integrated population in the upper band is plotted in \fref{fig:BI_A}(b). Here the trend follows closely the energy absorption as expected.

\subsection{Intermediate cases}
\label{subsec:IT}
\begin{figure*}[t!]\center
\includegraphics[scale=0.8]{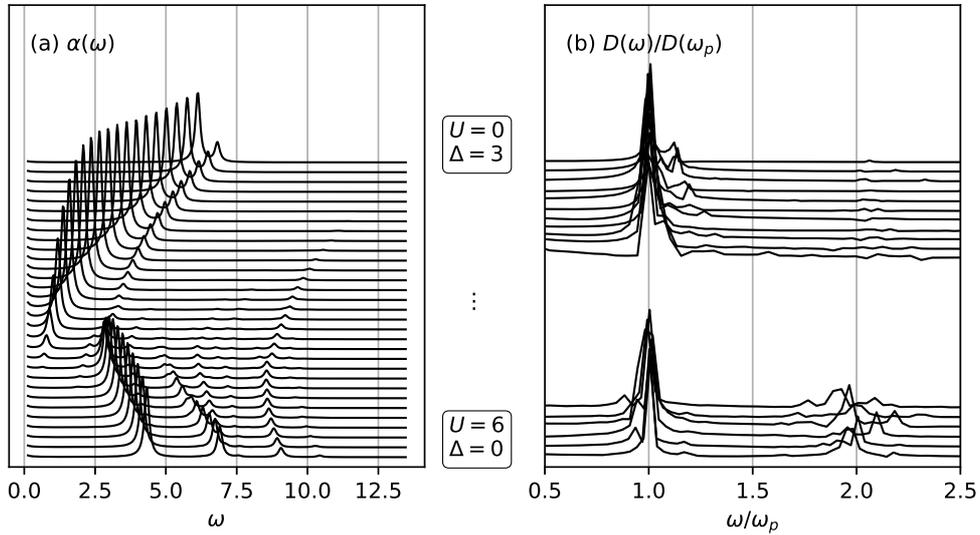}
\caption{(a) Absorption spectra for various $U$. (b) The Fourier spectra of Drude peaks $D(t)$ normalized by the $D(\omega_p)$. In both panels, lines are shifted by offsets depending on the corresponding parameters. The top line corresponds to $U=0, \Delta=3$, and the bottom one to $U=6, \Delta=3$. In between, the parameters change monotonically. In panel (b), data for $2.2 \leq U \leq 4.8$ are excluded, since there the system is initially metallic, and the optical excitation reduces the Drude peak.}
\label{fig:IT}
\end{figure*}

In order to investigate how the oscillations appearing in the Drude peak depend on the Hubbard interaction $U$ and the staggered potential $\Delta$, we consider intermediate cases here. For each parameter set, we set the pump frequency $\omega_p$ again at the first absorption peak and keep the pump strength $A_0 = 0.1$ with width $\tau = 5$ as before. We focus on the line $U + 2\Delta = 6$. In the thermodynamic limit, there exists a quantum phase transition between a MI and a BI around $U \simeq 2 \Delta \gg J_0$ \cite{kampf2003, manmana2004, otsuka2005}. In \fref{fig:IT}(a), we plot linear absorption spectra $\alpha(\omega)$ for different values of $U$. When $U=0$ (the BI case), there is only one peak at $\omega \approx 6.1$. This is due to the fact that the BI has two bands separated by $2\Delta$. As $U$ increases, the peak shifts towards lower energy. Near $U \approx 3.7$, the peak structure abruptly changes; the peak at $\omega \approx 0.9$ jumps to $\omega \approx 2.8$. This is related to a quantum phase transition, and due to the small system size, the critical value of $U$ seems shifted from the ideal one $U_c \approx 3$ to $3.7$. As $U$ further increases, additional peaks appear at higher energies. At $U=6$ (the MI case), peaks are located at $\omega \approx 4.3$, $6.9$, and $9.1$. When we pump at $\omega_p = 4.3$, we excite the states at $\omega_p$ and its higher harmonics. 

\Fref{fig:IT}(b) shows the Fourier spectra of the Drude peaks, $D(\omega)$, for various interaction strength $U$ and $\Delta$. For $2.2 \leq U \leq 4.8$, due to the closing of the energy gap, the system has initially a nonzero Drude peak, and the optical excitation reduces it. Since this is not the case that we are interested in, we exclude the data from the plot. We see that in the BI regime, there is no second harmonic peak, while in the MI regime, the second harmonic peak persists even with nonzero $\Delta$. This finding confirms that the second harmonic oscillation of the Drude peak is the signature of the correlation in the Mott insulator, and it evidently distinguishes between MI and BI regimes.

\section{Discussions}
\label{sec:diss}
\begin{figure*}[t!]\center
\includegraphics[scale=0.8]{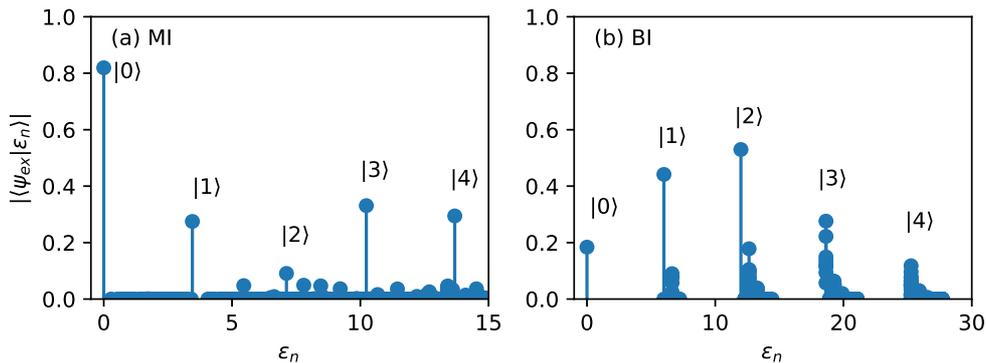}
\caption{The overlap of the excited state $\ket{\psi_{\rm ex}}$ with energy eigenstates $\ket{\epsilon_n}$ in the MI (left) and the BI (right).}
\label{fig:full_eig}
\end{figure*}

In this section, we elaborate on the properties of the photoexcited states by full-diagonalization of $L=8$ sites. All the energy eigenstates $\ket{\epsilon_n}$ for $n=1,\cdots,4900$ are calculated. We calculate excited states, $\ket{\psi_{\rm ex}}$, by applying photoexcitation of $A_0=0.4$ and $\tau = 5$ to the ground state. After the photoexcitation, the Hamiltonian becomes time-independent, and we can decompose the excited states by the eigenstates of the Hamiltonian $\ket{\epsilon_n}$. In \fref{fig:full_eig}, we plot the overlap of the photoexcited states with eigenstates, $| \braket{\psi_{\rm ex} | \epsilon_n}|$, in the MI and the BI. In both cases, there are non-negligible weights on the states whose energies are integer multiples of the pump frequency $\omega \approx k \omega_p$ ($k = 0, 1,2, \cdots$.). Here $k=1$ corresponds to the single photon absorption, which can be seen in the linear absorption spectrum. The states for $k \geq 2$ represent multi-photon excited (MPE) processes.  

\begin{figure*}[t!]\center
\includegraphics[scale=0.8]{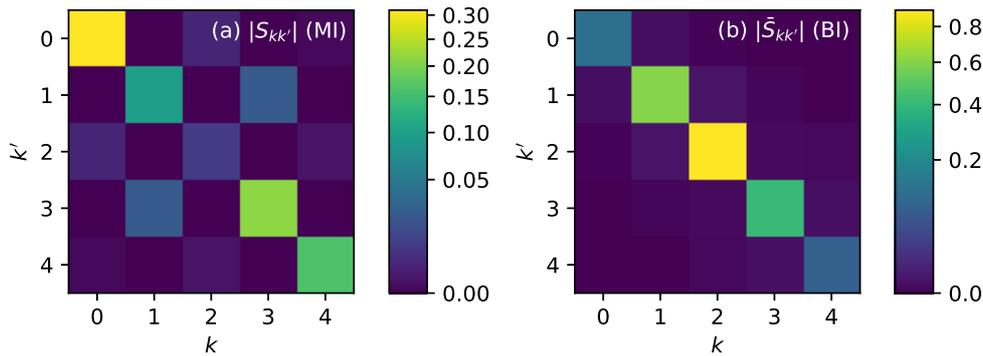}
\caption{The matrix element $S_{kk'}$ in \eref{eq:Skk} for the MI (left) and $\bar{S}_{kk'}$in \eref{eq:Skk2} for the BI (right).}
\label{fig:full_Skk}
\end{figure*}

Now we first consider the MI case. In the energy spectrum, there are many states between multiples of photon energies, $k\omega_p$, due to the correlation $U$. However, the weight is highly concentrated in the non-degenerate MPE states; for each $\omega \simeq k \omega_p$, there is only one dominant state. We label these states as $\ket{k}$ as in \fref{fig:full_eig}(a). The discreetness of the excited energy spectrum has been discussed as a result of charge-spin separation, which is a characteristic of one-dimensional systems \cite{mizuno2000, itoh2006, takahashi2008}. The contribution from the five states $\sqrt(\sum_{k=0}^4| \braket{\psi_{\rm ex} | k}|^2)$ accounts for more than $95\%$ of the total weight. Therefore, the excited states can be well described by
\begin{equation}
\ket{\psi_{\rm ex} (t)} \simeq \sum_{k=0}^{4} c_k e^{-i k \omega_pt }\ket{k}.
\end{equation}
Then expectation value of an operator $\hat{O}$ is approximated as 
\begin{equation}
\fl
\bra{\psi_{\rm ex} (t)} \hat{O} \ket{\psi_{\rm ex} (t)} \simeq \sum_{k,k' = 0}^4 e^{i(k-k')\omega_p t} c^*_k c_{k'} \braket{k|\hat{O}|k'} \equiv \sum_{k,k' = 0}^4 e^{i(k-k')\omega_p t} S_{kk'}.
\label{eq:Skk}
\end{equation}
The time-dependent oscillation comes from the off-diagonal elements, $k\neq k'$. In \fref{fig:full_Skk}(a), the matrix elements of $|S_{kk'}|$ for the double occupancy is plotted. There are non-negligible off-diagonal elements between $k=1$ and $k=3$, whose energy difference, $2\omega_p$, gives the second harmonic oscillations. Similar structures exist for the antiferromagnetic order parameter and the kinetic energy.

Next we turn to the BI. In the energy spectrum, the MPE states are separated by $2\Delta$, and there are no states in between. The MPE states are quasi-degenerate in contrast to the MI, where there is only one state for each $\omega \simeq k \omega_p$. Thus, the distribution of the weight is not as concentrated as in the MI case. The largest six states give about 80$\%$ of the total weight, and we need thirty six states to have more than 95$\%$. We label a state in the manifold of $k$-photon excited states as $\ket{m_k}$. The transition elements $S_{m_k, m'_{k'}}$ for the thirty six states are calculated. We find that there exist small amounts of off-diagonal elements between different manifolds, e.g., $k=2$ and $k=3$. However, we still need to sum over $m_k$ in \eref{eq:Skk}. Due to the near degeneracy of MPE states in each manifold $k$, we can approximate \eref{eq:Skk} as
\begin{equation}
\fl
\bra{\psi_{\rm ex} (t)} \hat{O} \ket{\psi_{\rm ex} (t)} \simeq \sum_{k,k' = 0}^4 e^{i(k-k')\omega_p t}  \sum_{m_k\in k, m'_{k'} \in k'} S_{m_k, m'_{k'}} \equiv \sum_{k,k' = 0}^4 e^{i(k-k')\omega_p t} \bar{S}_{kk'}.
\label{eq:Skk2}
\end{equation}
\Fref{fig:full_Skk}(b) shows the matrix elements of $|\bar{S}_{kk'}|$. In contrast to the MI case, the off-diagonal elements are ignorable. This is because the degenerate MPE states with different phases are summed up, and the constructive interference between different manifolds are destroyed. Thus, the second harmonic oscillation is absent.

\section{Conclusion}
\label{sec:conc}
In this work, a photoexcited one-dimensional ionic Hubbard model is studied by an exact diagonalization method. We find that the energy absorption in Mott and band insulators exhibits nonlinear dependence on the pump amplitude. A nonzero Drude peak in conductivity or an in-gap density of states appear after short optical excitation indicating metallic behaviors, which can be measured by time-resolved experiments. In a Mott insulating regime, the Drude peak oscillates at the pump frequency and its second harmonic. The former originates from the current oscillation, and the latter from an interference between single- and three-photon excited states. We expect that as far as strong correlation exists, such coherent oscillations (may not be only the second harmonic) appear in other one-dimensional systems. Coherent oscillations also appear in the double occupancy and the kinetic energy, which might be easier to detect experimentally. In the band insulating regime, the Drude peak oscillates only at the pump frequency. This is because the superposition of the degenerate multi-photon excited states with different phases destroy the constructive interference. The relation between the Drude weight and the in-gap density of states are also discussed. Including dynamical phonons is a next move to account for realistic experimental results \cite{yonemitsu2004, matsubara2014}. Time-dependent spectroscopies on systems with modulated hopping \cite{maeshima2006, atala2013, rincon2014a} or multi-orbitals \cite{tanaka2018, rincon2018a} are also interesting open problems.

\ack
We thank S. Stumper, S. A. Sato, M. Thoss and M. \v{Z}onda for useful discussions. This work is supported by Research Foundation for Opto-Science and Technology and by Georg H. Endress Foundation. The author acknowledges support by the state of Baden-W\"urttemberg through bwHPC.

\appendix
\section{Transient behavior of single-particle spectral functions}
\label{sec:AppA}
\begin{figure*}[t!]\center
\includegraphics[scale=0.8]{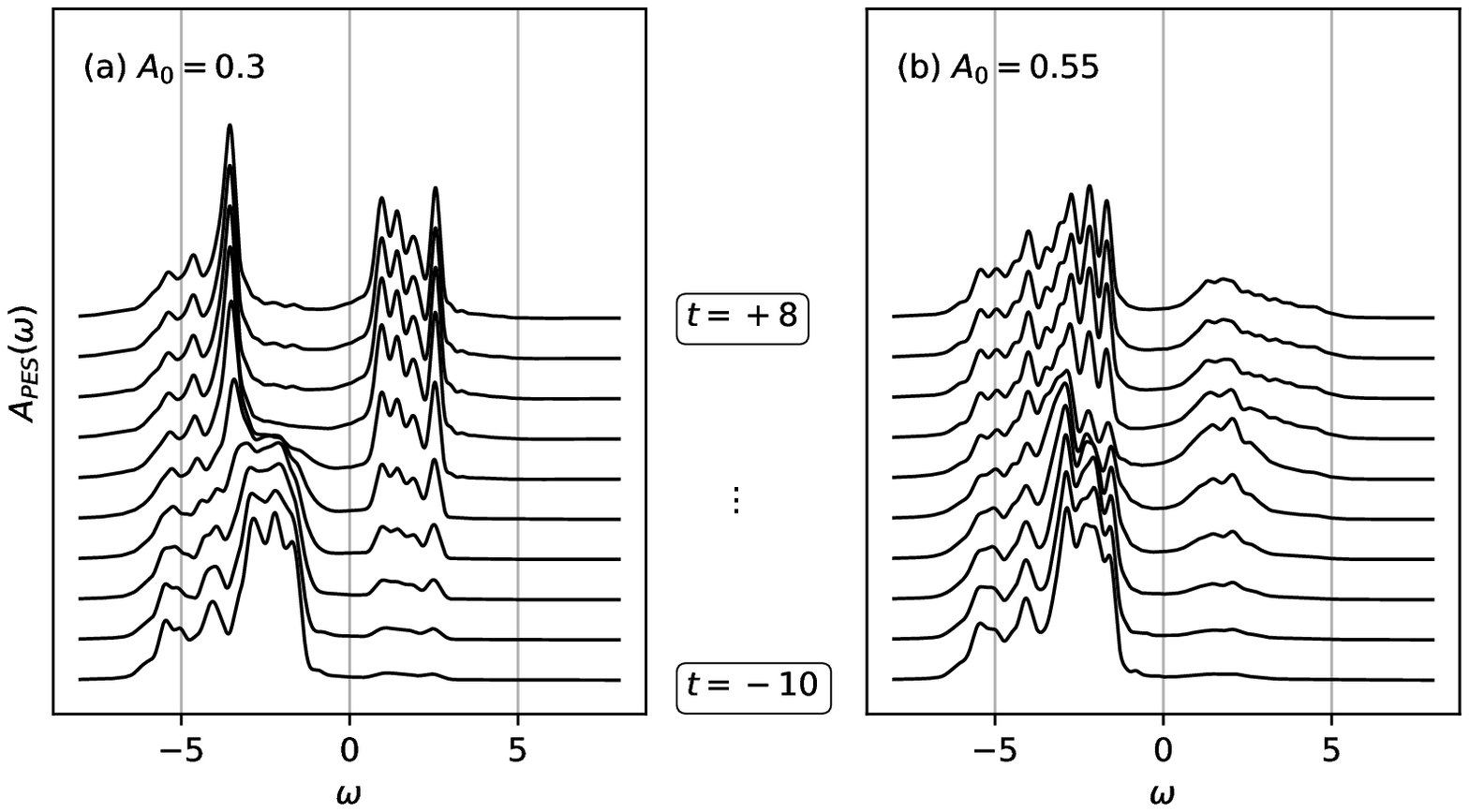}
\caption{The transient photoemission spectra for the MI for $A_0 = 0.3$ (left) and $A_0 = 0.55$ (right). The lines are shifted by offsets corresponding to the pump-probe delay time $t$. }
\label{fig:AppA}
\end{figure*}

In \sref{sec:results}, the single-particle spectral functions at $t_f = 50$ are shown. Here we show the transient behaviors of the photoemission spectra of the MI, $A_{\rm PES} (t, \omega)$, for $A_0= 0.3$ and $0.55$ in \fref{fig:AppA}. We see that the change of the spectral functions occurs within the pump-pulse duration $-\tau < t < \tau$ with $\tau = 5$. Also, due to the relatively wide time convolution, $2T = 20$, in \eref{eq:A_PES}, the temporal change is rather smooth; in both cases, the weight in the lower Hubbard band is gradually transferred to the upper Hubbard band. For $A_0=0.3$, the weight in the lower Hubbard band $-3.5 \lesssim \omega \lesssim 1.5$ is depleted, and the shape of the spectral  distribution in the upper Hubbard band does not change much. This resembles the quantum tunneling regime \cite{oka2012}, while the precise connection is unclear at this moment. On the other hand, for $A_0 = 0.5$, the distribution is first accumulated around the bottom of the upper Hubbard band, and then becomes wider at later times.

\section{Antiferromagnetic order}
\label{sec:AppB}
\begin{figure*}[t!]\center
\includegraphics[scale=0.8]{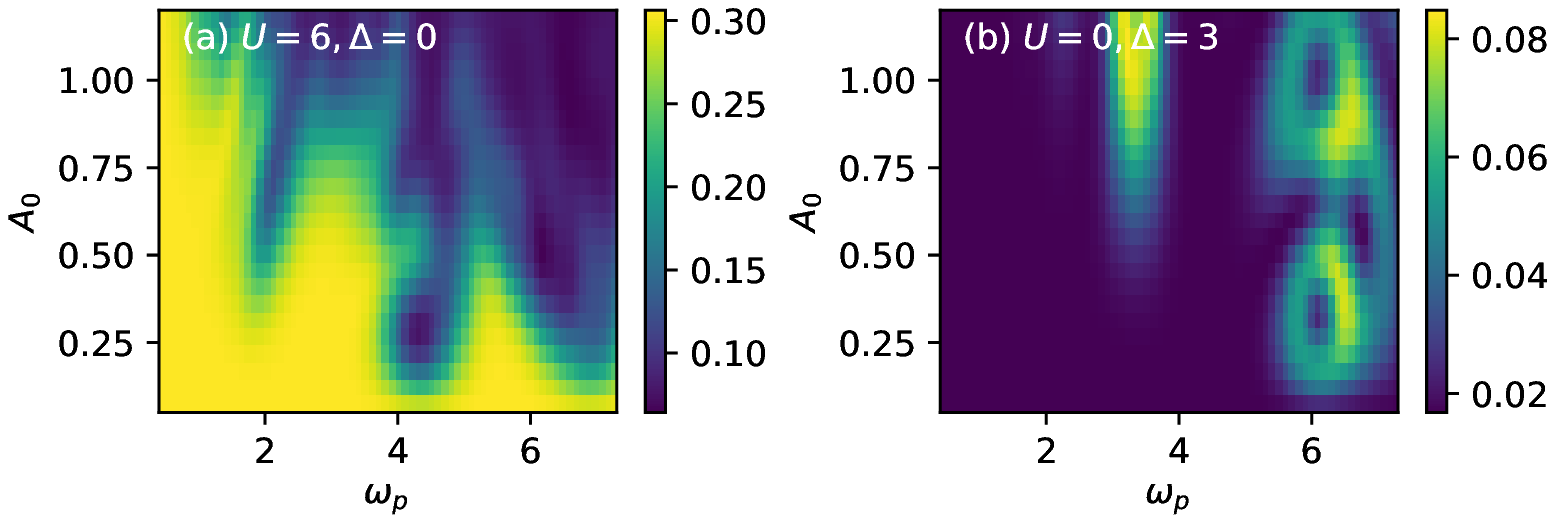}
\caption{Antiferromagnetic order after photoexcitation of strength $A_0$ and frequency $\omega_p$ in the MI (left) or the BI (right).}
\label{fig:AppB_SDW}
\end{figure*}
We show antiferromagnetic (AFM) order after optical excitation of strength $A_0$ and frequency $\omega_p$ in \fref{fig:AppB_SDW} for the MI and the BI cases. In the MI, there exists AFM order initially, and the optical excitation destroys it. The amount of reduction closely correlates with the amount of energy absorption in \fref{fig:MI_scan}. On the other hand, in the BI, there is no AFM order in the ground state. Optical excitation then induces slight AFM order near the energy absorption peaks, $\omega_p = 6.1$ and $6.9$. However, the detailed structures near these frequencies in \fref{fig:AppB_SDW}(b) do not follow that of \fref{fig:BI_scan}. In particular, at $\omega_p=6.1$, the energy absorption shows Rabi-like oscillations in terms of $A_0$, whose period is $\Delta A_0 \approx 0.6$, while that of the AFM order is roughly half, $\Delta A_0 \approx 0.3$. Similarly, at $\omega_p=6.9$, the period of the oscillation is $\Delta A_0 \approx 1.2$ in the energy absorption, but $\Delta A_0 \approx 0.6$ in the AFM order. The discrepancy of these frequencies in the energy and the AFM order also appears in terms of the pump duration $\tau$; again the period of the AFM order is half of the period of the energy. Since the Rabi frequency is proportional to the coupling constant between the system and the electromagnetic fields, the above observation indicates that the coupling constant of the AFM channel is twice bigger than that of the energy channel in the BI.

\section{Evaluation of \eref{eq:A_formula} for noninteracting systems}
\label{sec:AppC}
In this section, we evaluate the expression \eref{eq:A_formula} for a noninteracting and time-independent Hamiltonian $H_0 = \sum_{k} \xi_{k} c^{\dagger}_{k} c_{k}$ for an arbitrary initial state. The eigenstates of the Hamiltonian are simple Fock states
\begin{equation}
\ket{m} = c^{\dagger}_{k^m_1} c^{\dagger}_{k^m_2} \cdots c^{\dagger}_{k^m_N} \ket{0},
\end{equation}
where $\ket{0}$ is a vacuum state. The corresponding eigenenergy is $\epsilon_m = \sum_{l=1}^{N} \xi_{k^m_l}$. An initial state can be decomposed by the Fock states as
\begin{equation}
\ket{\Phi_0} = \sum_{m} p_m \ket{m}.
\end{equation}
Now we consider $A^{\alpha}_{\rm PES} (t, \omega)$ for an operator $c_{\alpha} = \sum_{k} a_{\alpha k} c_{k}$. \eref{eq:A_PES} becomes 
\begin{equation}
\fl \int_{-T}^{T} d \tau_{1} \int_{-T}^{T} d \tau_{2}e^{-i \omega\left(\tau_{1}-\tau_{2}\right)} \sum_{m,n, k, k'} p_m p_n^* a_{\alpha k} a_{\alpha k'}^* \bra{n} c^{\dagger}_{k'} c_{k}\ket{m} e^{i (\tau_1 \xi_{k'} - \tau_2 \xi_k)} e^{i t (\xi_k -\xi_{k'})}.
\end{equation}
For large $t$, only $k=k'$ contribution remains in the last exponential factor. Thus, the expression is reduced to
\begin{equation}
\sum_{m,k} |p_m|^2 |a_{\alpha k}|^2 \bra{m} n_k \ket{m} \frac{\sin^2 \left[ T (\omega - \xi_k) \right]}{T^2 (\omega - \xi_k)^2},
\end{equation}
which gives a peak at $\xi_k$ of width $\pi/T$ and of height determined by the occupation of single-particle state $k$. The inverse photoemission part $A^{\alpha}_{\rm IPES} (t, \omega)$ gives a similar expression 
\begin{equation}
\sum_{m,k} |p_m|^2 |a_{\alpha k}|^2 \bra{m} (1-n_k) \ket{m} \frac{\sin^2 \left[ T (\omega - \xi_k) \right]}{T^2 (\omega - \xi_k)^2}.
\end{equation}
In the end, the sum of the two contributions is reduced to
\begin{equation}
A^{\alpha} (\omega) = \sum_{k} |a_{\alpha k}|^2 \frac{\sin^2 \left[ T (\omega - \xi_k) \right]}{T^2 (\omega - \xi_k)^2},
\end{equation}
which is independent of the initial distribution $\{ p_m \}$. This is similar to the case in equilibrium, where $A^{\alpha}(\omega)$ is independent of temperatures for noninteracting systems \cite{bruus2004}.

\


\begin{thebibliography}{10}
\expandafter\ifx\csname url\endcsname\relax
  \def\url#1{{\tt #1}}\fi
\expandafter\ifx\csname urlprefix\endcsname\relax\def\urlprefix{URL }\fi
\providecommand{\eprint}[2][]{\url{#2}}

\bibitem{pitaevskii1981}
Pitaevskii L~P and Lifshitz E~M 1981 {\em Physical {{Kinetics}}\/} ({Oxford}:
  {Butterworth-Heinemann})

\bibitem{carmelo2007}
Carmelo J~M~P, {Lopes dos Santos} J~M~B, Rocha~Vieira V and Sacramento P 2007
  {\em Strongly {{Correlated Systems}}, {{Coherence And Entanglement}}\/}
  ({Singapore}: {World Scientific})

\bibitem{haug2008}
Haug H and Jauho A~P 2008 {\em Quantum {{Kinetics}} in {{Transport}} and
  {{Optics}} of {{Semiconductors}}\/} ({Berlin; Heidelberg}: {Springer-Verlag})

\bibitem{yamada2010}
Yamada K 2010 {\em Electron {{Correlation}} in {{Metals}}\/} ({Cambridge}:
  {Cambridge University Press})

\bibitem{avella2011}
Avella A and Mancini F 2011 {\em Strongly {{Correlated Systems}}: {{Theoretical
  Methods}}\/} ({Berlin; Heidelberg}: {Springer-Verlag})

\bibitem{avella2013}
Avella A and Mancini F 2013 {\em Strongly {{Correlated Systems}}: {{Numerical
  Methods}}\/} ({Berlin; Heidelberg}: {Springer-Verlag})

\bibitem{avella2016}
Avella A and Mancini F 2016 {\em Strongly {{Correlated Systems: Experimental
  Techniques}}\/} ({Berlin; Heidelberg}: {Springer-Verlag})

\bibitem{averitt2002}
Averitt R~D and Taylor A~J 2002 {\em J. Phys. Condens. Matter\/} {\bf 14} R1357

\bibitem{yonemitsu2008}
Yonemitsu K and Nasu K 2008 {\em Phys. Rep.\/} {\bf 465} 1

\bibitem{giannetti2016}
Giannetti C, Capone M, Fausti D, Fabrizio M, Parmigiani F and Mihailovic D 2016
  {\em Adv. Phys.\/} {\bf 65} 58

\bibitem{nicoletti2016}
Nicoletti D and Cavalleri A 2016 {\em Adv. Opt. Photon.\/} {\bf 8} 401

\bibitem{citro2018}
Citro R and Mancini F 2018 {\em Out-of-{{Equilibrium Physics}} of {{Correlated
  Electron Systems}}\/} ({Cham}: {Springer International Publishing})

\bibitem{ishihara2019}
Ishihara S 2019 {\em J. Phys. Soc. Jpn.\/} {\bf 88} 072001

\bibitem{schmuttenmaer2004}
Schmuttenmaer C~A 2004 {\em Chem. Rev.\/} {\bf 104} 1759

\bibitem{zhou2018}
Zhou X, He S, Liu G, Zhao L, Yu L and Zhang W 2018 {\em Rep. Prog. Phys.\/}
  {\bf 81} 062101

\bibitem{lv2019}
Lv B, Qian T and Ding H 2019 {\em Nat. Rev. Phys.\/} {\bf 1} 609

\bibitem{wark1996}
Wark J 1996 {\em Contemp. Phys.\/} {\bf 37} 205

\bibitem{buzzimichele2019}
{Buzzi Michele}, {F{\"o}rst Michael} and {Cavalleri Andrea} 2019 {\em Philos.
  Trans. R. Soc. Math. Phys. Eng. Sci.\/} {\bf 377} 20170478

\bibitem{fausti2011}
Fausti D, Tobey R~I, Dean N, Kaiser S, Dienst A, Hoffmann M~C, Pyon S, Takayama
  T, Takagi H and Cavalleri A 2011 {\em Science\/} {\bf 331} 189

\bibitem{hu2014}
Hu W, Kaiser S, Nicoletti D, Hunt C~R, Gierz I, Hoffmann M~C, Le~Tacon M, Loew
  T, Keimer B and Cavalleri A 2014 {\em Nat. Mater.\/} {\bf 13} 705

\bibitem{mitrano2016}
Mitrano M, Cantaluppi A, Nicoletti D, Kaiser S, Perucchi A, Lupi S, Di~Pietro
  P, Pontiroli D, Ricc{\`o} M, Clark S~R, Jaksch D and Cavalleri A 2016 {\em
  Nature\/} {\bf 530} 461

\bibitem{stojchevska2014}
Stojchevska L, Vaskivskyi I, Mertelj T, Kusar P, Svetin D, Brazovskii S and
  Mihailovic D 2014 {\em Science\/} {\bf 344} 177

\bibitem{rini2007}
Rini M, Tobey R, Dean N, Itatani J, Tomioka Y, Tokura Y, Schoenlein R~W and
  Cavalleri A 2007 {\em Nature\/} {\bf 449} 72

\bibitem{tobey2008}
Tobey R~I, Prabhakaran D, Boothroyd A~T and Cavalleri A 2008 {\em Phys. Rev.
  Lett.\/} {\bf 101} 197404

\bibitem{Aoki2014}
Aoki H, Tsuji N, Eckstein M, Kollar M, Oka T and Werner P 2014 {\em Rev. Mod.
  Phys.\/} {\bf 86} 779

\bibitem{cazalilla2002}
Cazalilla M~A and Marston J~B 2002 {\em Phys. Rev. Lett.\/} {\bf 88} 256403

\bibitem{vidal2004}
Vidal G 2004 {\em Phys. Rev. Lett.\/} {\bf 93} 040502

\bibitem{daley2004}
Daley A~J, Kollath C, Schollw{\"o}ck U and Vidal G 2004 {\em J. Stat. Mech.:
  Theor. Exp.\/} {\bf 2004} P04005

\bibitem{white2004}
White S~R and Feiguin A~E 2004 {\em Phys. Rev. Lett.\/} {\bf 93} 076401

\bibitem{garcia-ripoll2006}
{Garc{\'i}a-Ripoll} J~J 2006 {\em New J. Phys.\/} {\bf 8} 305

\bibitem{paeckel2019}
Paeckel S, K{\"o}hler T, Swoboda A, Manmana S~R, Schollw{\"o}ck U and Hubig C
  2019 {\em Annals of Physics\/} {\bf 411} 167998

\bibitem{schiro2010}
Schir{\'o} M and Fabrizio M 2010 {\em Phys. Rev. Lett.\/} {\bf 105} 076401

\bibitem{fabrizio2013}
Fabrizio M 2013 The {{Out}}-of-{{Equilibrium Time}}-{{Dependent Gutzwiller
  Approximation}} {\em New {{Materials}} for {{Thermoelectric Applications}}:
  {{Theory}} and {{Experiment}}\/} ({Dordrecht}: {Springer Netherlands}) p 247

\bibitem{goth2012}
Goth F and Assaad F~F 2012 {\em Phys. Rev. B\/} {\bf 85} 085129

\bibitem{ido2015}
Ido K, Ohgoe T and Imada M 2015 {\em Phys. Rev. B\/} {\bf 92} 245106

\bibitem{park1986}
Park T~J and Light J~C 1986 {\em J. Chem. Phys.\/} {\bf 85} 5870

\bibitem{manmana2005}
Manmana S~R, Muramatsu A and Noack R~M 2005 {\em AIP Conf. Proc.\/} {\bf 789}
  269

\bibitem{balzer2012}
Balzer M, Gdaniec N and Potthoff M 2012 {\em J. Phys. Condens. Matter\/} {\bf
  24} 035603

\bibitem{eckstein2009}
Eckstein M, Hackl A, Kehrein S, Kollar M, Moeckel M, Werner P and Wolf F~A 2009
  {\em Eur. Phys. J. Spec. Top.\/} {\bf 180} 217

\bibitem{Orus2014}
Or{\'u}s R 2014 {\em Annals of Physics\/} {\bf 349} 117

\bibitem{falicov1969}
Falicov L~M and Kimball J~C 1969 {\em Phys. Rev. Lett.\/} {\bf 22} 997

\bibitem{atala2013}
Atala M, Aidelsburger M, Barreiro J~T, Abanin D, Kitagawa T, Demler E and Bloch
  I 2013 {\em Nature Phys\/} {\bf 9} 795

\bibitem{lenarcic2014}
Lenar{\v c}i{\v c} Z, Gole{\v z} D, Bon{\v c}a J and Prelov{\v s}ek P 2014 {\em
  Phys. Rev. B\/} {\bf 89} 125123

\bibitem{lu2015}
Lu H, Shao C, Bon{\v c}a J, Manske D and Tohyama T 2015 {\em Phys. Rev. B\/}
  {\bf 91} 245117

\bibitem{shao2016}
Shao C, Tohyama T, Luo H~G and Lu H 2016 {\em Phys. Rev. B\/} {\bf 93} 195144

\bibitem{wang2017}
Wang Y, Claassen M, Moritz B and Devereaux T~P 2017 {\em Phys. Rev. B\/} {\bf
  96} 235142

\bibitem{wang2018}
Wang Y, Chen C~C, Moritz B and Devereaux T~P 2018 {\em Phys. Rev. Lett.\/} {\bf
  120} 246402

\bibitem{nagaosa1986}
Nagaosa N and Takimoto J~i 1986 {\em J. Phys. Soc. Jpn.\/} {\bf 55} 2735

\bibitem{egami1993}
Egami T, Ishihara S and Tachiki M 1993 {\em Science\/} {\bf 261} 1307

\bibitem{ishihara1994}
Ishihara S, Egami T and Tachiki M 1994 {\em Phys. Rev. B\/} {\bf 49} 8944

\bibitem{maeshima2005}
Maeshima N and Yonemitsu K 2005 {\em J. Phys. Conf. Ser.\/} {\bf 21} 183

\bibitem{resta1995}
Resta R and Sorella S 1995 {\em Phys. Rev. Lett.\/} {\bf 74} 4738

\bibitem{valiente2010}
Valiente M, K{\"u}ster M and Saenz A 2010 {\em EPL\/} {\bf 92} 10001

\bibitem{lanczos1950}
Lanczos C 1950 {\em J. Res. Natl. Bur. Stand.\/} {\bf 45} 255

\bibitem{kindt1999}
Kindt J~T and Schmuttenmaer C~A 1999 {\em J. Chem. Phys.\/} {\bf 110} 8589

\bibitem{nemec2002}
N{\v e}mec H, Kadlec F and Ku{\v z}el P 2002 {\em J. Chem. Phys.\/} {\bf 117}
  8454

\bibitem{orenstein2015}
Orenstein J and Dodge J~S 2015 {\em Phys. Rev. B\/} {\bf 92} 134507

\bibitem{kennes2017}
Kennes D~M, Wilner E~Y, Reichman D~R and Millis A~J 2017 {\em Phys. Rev. B\/}
  {\bf 96} 054506

\bibitem{okamoto2017}
Okamoto J~i, Hu W, Cavalleri A and Mathey L 2017 {\em Phys. Rev. B\/} {\bf 96}
  144505

\bibitem{vengurlekar1988}
Vengurlekar A~S and Jha S~S 1988 {\em Phys. Rev. B\/} {\bf 38} 2044

\bibitem{freericks2009}
Freericks J~K, Krishnamurthy H~R and Pruschke T 2009 {\em Phys. Rev. Lett.\/}
  {\bf 102} 136401

\bibitem{braun2015}
Braun J, Rausch R, Potthoff M, Min{\'a}r J and Ebert H 2015 {\em Phys. Rev.
  B\/} {\bf 91} 035119

\bibitem{sentef2017}
Sentef M~A 2017 {\em Phys. Rev. B\/} {\bf 95} 205111

\bibitem{acciai2019}
Acciai M, Calzona A, Carrega M, Martin T and Sassetti M 2019 {\em New J.
  Phys.\/} {\bf 21} 103031

\bibitem{dagotto1994}
Dagotto E 1994 {\em Rev. Mod. Phys.\/} {\bf 66} 763

\bibitem{matsueda2007}
Matsueda H and Ishihara S 2007 {\em J. Phys. Soc. Jpn.\/} {\bf 76} 083703

\bibitem{lu2012}
Lu H, Sota S, Matsueda H, Bon{\v c}a J and Tohyama T 2012 {\em Phys. Rev.
  Lett.\/} {\bf 109} 197401

\bibitem{maeshima2005a}
Maeshima N and Yonemitsu K 2005 {\em J. Phys. Soc. Jpn.\/} {\bf 74} 2671

\bibitem{kampf2003}
Kampf A~P, Sekania M, Japaridze G~I and Brune P 2003 {\em J. Phys. Condens.
  Matter\/} {\bf 15} 5895

\bibitem{manmana2004}
Manmana S~R, Meden V, Noack R~M and Sch{\"o}nhammer K 2004 {\em Phys. Rev. B\/}
  {\bf 70} 155115

\bibitem{otsuka2005}
Otsuka H and Nakamura M 2005 {\em Phys. Rev. B\/} {\bf 71} 155105

\bibitem{mizuno2000}
Mizuno Y, Tsutsui K, Tohyama T and Maekawa S 2000 {\em Phys. Rev. B\/} {\bf 62}
  R4769

\bibitem{itoh2006}
Itoh H, Takahashi A and Aihara M 2006 {\em Phys. Rev. B\/} {\bf 73} 075110

\bibitem{takahashi2008}
Takahashi A, Itoh H and Aihara M 2008 {\em Phys. Rev. B\/} {\bf 77} 205105

\bibitem{yonemitsu2004}
Yonemitsu K 2004 {\em J. Phys. Soc. Jpn.\/} {\bf 73} 2868

\bibitem{matsubara2014}
Matsubara Y, Ogihara S, Itatani J, Maeshima N, Yonemitsu K, Ishikawa T, Okimoto
  Y, Koshihara S~y, Hiramatsu T, Nakano Y, Yamochi H, Saito G and Onda K 2014
  {\em Phys. Rev. B\/} {\bf 89} 161102

\bibitem{maeshima2006}
Maeshima N and Yonemitsu K 2006 {\em Phys. Rev. B\/} {\bf 74} 155105

\bibitem{rincon2014a}
Rincon J, {Al-Hassanieh} K~A, Feiguin A~E and Dagotto E 2014 {\em Phys. Rev.
  B\/} {\bf 90} 155112

\bibitem{tanaka2018}
Tanaka Y, Daira M and Yonemitsu K 2018 {\em Phys. Rev. B\/} {\bf 97} 115105

\bibitem{rincon2018a}
Rincon J, Dagotto E and Feiguin A~E 2018 {\em Phys. Rev. B\/} {\bf 97} 235104

\bibitem{oka2012}
Oka T 2012 {\em Phys. Rev. B\/} {\bf 86} 075148

\bibitem{bruus2004}
Bruus H and Flensberg K 2004 {\em Many-{{Body Quantum Theory}} in {{Condensed
  Matter Physics}}: {{An Introduction}}\/} ({Oxford}: {Oxford University
  Press})

\end{thebibliography}
\providecommand{\newblock}{}

\end{document}